*Research Article*

# Modulation of the Cardiomyocyte Contraction inside a Hydrostatic Pressure Bioreactor: *In Vitro* Verification of the Frank-Starling Law

**Lorenzo Fassina,**[1,2] **Giovanni Magenes,**[1,2] **Roberto Gimmelli,**[3] **and Fabio Naro**[4]

[1]*Dipartimento di Ingegneria Industriale e dell'Informazione, Università di Pavia, Via Ferrata 1, 27100 Pavia, Italy*
[2]*Centro di Ingegneria Tissutale (CIT), Università di Pavia, 27100 Pavia, Italy*
[3]*Dipartimento di Medicina Sperimentale, Università "Sapienza," 00161 Roma, Italy*
[4]*Dipartimento di Scienze Anatomiche, Istologiche, Medico-Legali e dell'Apparato Locomotore, Università "Sapienza," 00161 Roma, Italy*

Correspondence should be addressed to Lorenzo Fassina; lorenzo.fassina@unipv.it

We have studied beating mouse cardiac syncytia *in vitro* in order to assess the inotropic, ergotropic, and chronotropic effects of both increasing and decreasing hydrostatic pressures. In particular, we have performed an image processing analysis to evaluate the kinematics and the dynamics of those pressure-loaded beating syncytia starting from the video registration of their contraction movement. By this analysis, we have verified the Frank-Starling law of the heart in *in vitro* beating cardiac syncytia and we have obtained their geometrical-functional classification.

## 1. Introduction

Understanding how cells react to mechanical forces is crucial. For instance, when osteoblasts sense a fluid shear stress, stretch-gated ion channels open and specific intracellular mechanisms lead to an enhanced production of bone matrix [1–3]. On the other hand, both tension and compression modulate the expression of transcription factors essential for the homeostasis of bone, cartilage, and tooth tissues [4]. Compression has a role during embryogenesis, when the blastocoel fluid presses the inner cell mass and, as a consequence, activates the key transcription factors OCT4, SOX2, and NANOG that determine pluripotency in the epiblast [5–7]. Tension and compression may also change the transcription more rapidly when they are transmitted directly into the nucleus via the cytoskeleton linked to nuclear envelope proteins [8].

The preceding examples of structure/force/function relationships are well understandable in the frame of the "tensegrity" theory [9–13]: the functions of cells and tissues can be modulated not only by molecules, but also by biophysical stimuli. In particular, during the *in vitro* culture inside bioreactors, the biophysical forces may modify a specific cell status of force equilibrium, named "tensegrity," inducing, via mechanotransduction, changes to the transcriptional profile. In general, the mechanical bioreactors elicit time varying forces acting perpendicularly or tangentially onto the cells and modulating the cell tensegrity via tensile, compressive, and shear deformations [14, 15].

In the present study, an instantaneous modulation of the cell function (without the involvement of transcriptional mechanisms) is well exemplified by the cardiomyocytes subjected to mechanical forces according to the Frank-Starling law of the heart [16].

In a previous work, to extend the possible use of beating cardiac syncytia cultured *in vitro* (e.g., in studies about human cardiac syncytium in physiological and pathological conditions, patient-tailored therapeutics, and syncytium models derived from induced pluripotent/embryonic stem cells with genetic mutations), we have developed a novel method based on image processing analysis to evaluate the kinematics and the dynamics of *in vitro* beating syncytia starting from the video registration of their contraction movement [17]: in particular, our method uses the displacement vector field and the velocity vector field of a beating patch to evaluate the syncytium not only from the chronotropic viewpoint, but also from the inotropic and ergotropic ones.

In the present work, the preceding calculus method allowed the study of the mechanical modulation of the contraction properties in *in vitro* beating cardiac syncytia as they were loaded with different hydrostatic pressures inside a bioreactor. In particular, the computed kinematic and dynamic parameters aimed at revealing the inotropic, ergotropic, and chronotropic effects of the applied hydrostatic pressures: we anticipate that the data analysis permitted the verification of the Frank-Starling law in *in vitro* beating cardiac syncytia and their geometrical-functional classification.

## 2. Materials and Methods

*2.1. Beating Mouse Cardiac Syncytia.* Spontaneously beating cardiac syncytia were obtained from hearts of 1- to 2-day-old CD-1 mouse pups (Charles River Laboratories Italia, Calco, Italy), as previously described [18–20] with some modifications. Briefly, beating primary cultures of murine cardiomyocytes were prepared *in vitro* as follows: the hearts were quickly excised, the atria were cut off, and the ventricles were minced and digested by incubation with 100 $\mu$g/mL type II collagenase (Invitrogen, Carlsbad, CA) in ADS buffer (0.1 M HEPES, 0.1 M D-glucose, 0.5 M NaCl, 0.1 M KCl, 0.1 M $NaH_2PO_4 \cdot H_2O$, and 0.1 M $MgSO_4$) for 15 min at 37°C and then by incubation with 900 $\mu$g/mL pancreatin (Sigma-Aldrich, Milan, Italy) in ADS buffer for 15 min at 37°C. The resulting cell suspension was preplated for 2 h at 37°C to reduce the contribution of nonmyocardial cells. The unattached, cardiomyocyte-enriched cells remaining in suspension were collected, plated onto collagen-coated 35 mm Petri dishes, and covered by DMEM containing 10% horse serum, 5% fetal bovine serum, and 1× gentamicin (Roche Molecular Biochemicals, Indianapolis, IN). About $3 \times 10^5$ cardiomyocytes were cultured in each Petri dish at 37°C and 5% $CO_2$ to form a spontaneously beating cardiac syncytium.

*2.2. Pressure Bioreactor.* On day 3 of culture, at a constant temperature of 37°C and 5% $CO_2$, each syncytium was observed via a movie capture system (ProgRes C5, Jenoptik, Germany) in four different conditions of relative hydrostatic pressure: at 0 (atmospheric pressure), 100, 200, and 300 mmHg inside a custom-machined polymethylmethacrylate pressure chamber (Figure 1).

The syncytium cultures were repeated 4 times for a total of 40 syncytia loaded by both increasing and decreasing pressures. In particular, for each syncytium, we have obtained a sequence of eight video files in AVI format, each video with a duration of 40 s: at 0, 100, 200, and 300 mmHg (after that there was a recovery of 0.5 h in incubator at 37°C and 5% $CO_2$) and then at 300, 200, 100, and 0 mmHg.

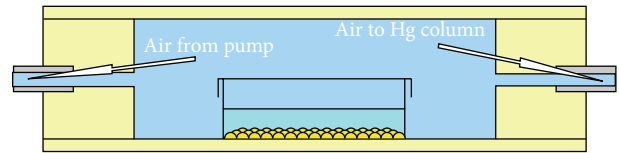

Figure 1: Scheme of the transversal section of the pressure chamber with a seeded Petri dish inside. Each syncytium was observed via a movie capture system in four different conditions of relative hydrostatic pressure: at 0 (atmospheric pressure), 100, 200, and 300 mmHg inside a custom-machined polymethylmethacrylate pressure chamber.

*2.3. Registration of the Syncytium Movement via the Apposition of Software Markers.* By the Video Spot Tracker (VST) program, which is used to track the motion of one or more spots in an AVI video file (http://cismm.cs.unc.edu/downloads/), in each video, we have systematically selected 30 spots or markers onto the first video frame, according to the same orthogonal grid. By starting the videos in VST, frame by frame, the program followed and registered the spatial-temporal coordinates $x$, $y$, and $t$ for each marker, as previously described [17]. The coordinates $x$ and $y$ are expressed in [pixel], whereas the coordinate $t$ is expressed in [s].

*2.4. Kinematics and Dynamics of the Beating Syncytium.* By an algorithm based on the Matlab programming language (The MathWorks, Inc., Natick, MA), frame by frame and for each marker, we have studied the kinematics and the dynamics of the beating cardiac syncytia, as previously described [17] (see the Appendix below for the mathematical details). In particular, in this work, we have evaluated the syncytium contraction in terms of contractility (maximum contraction velocity), contraction force, kinetic energy, and beat frequency.

*2.5. Statistics.* In order to compare the results between the different pressure conditions, one-way analysis of variance (ANOVA) with *post hoc* least significance difference (LSD) test was applied, electing a significance level of 0.05. The results are expressed as mean ± standard deviation.

## 3. Results

The software and calculus method previously described allowed the study of the mechanical modulation of the contraction properties in *in vitro* beating cardiac syncytia as they were loaded with different hydrostatic pressures. In particular, the computed kinematic and dynamic parameters aimed at revealing possible inotropic, ergotropic, and chronotropic effects of the applied hydrostatic pressures.

*3.1. Geometrical-Functional Classification of the Beating Cardiac Syncytia.* We observed particularly thick multilayers with "spheroidal" shape and "flat" multilayers (Figure 2), and we evaluated them in terms of contractility, contraction force, kinetic energy, and beat frequency.

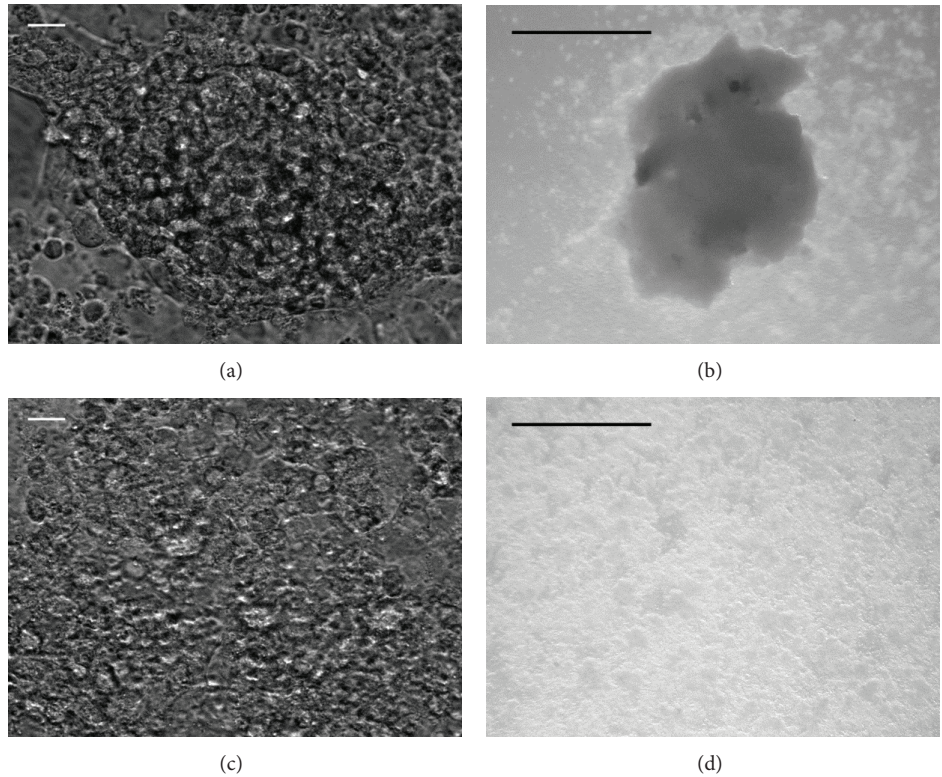

Figure 2: Cellular layers in beating syncytia. We observed particularly thick multilayers with "spheroidal" shape ((a) and (b), with bars equal to 50 µm and to 1 mm, resp.) and "flat" multilayers ((c) and (d), with bars equal to 50 µm and to 1 mm, resp.).

The spheroidal syncytia showed minimum contractility, minimum contraction force, minimum kinetic energy, and maximum beat frequency at 200 mmHg, with $P < 0.05$ in the comparison with the corresponding adjacent values at 100 and 300 mmHg (Figures 3, 4, and 6), except for the kinetic energy ($P > 0.05$) (Figure 5).

The flat syncytia were characterized by maximum contractility, maximum contraction force, maximum kinetic energy, and minimum beat frequency at 100 mmHg, with $P < 0.05$ in the comparison with the corresponding adjacent values at 0 and 200 mmHg (Figures 4, 5, and 6), except for the contractility ($P > 0.05$) (Figure 3).

At all pressures, the contractility, the contraction force, and the kinetic energy of spheroidal syncytia were significantly higher than those of flat syncytia ($P < 0.05$) (Figures 3, 4, and 5), except for the contractility at 200 mmHg ($P > 0.05$) (Figure 3). In addition, at all pressures, the beat frequency of spheroidal syncytia was significantly lower than the beat frequency of flat syncytia ($P < 0.05$) (Figure 6).

## 4. Discussion

The mouse has become a cornerstone of the heart research because of the high potential in manipulating its genome and the consequent availability of models of cardiovascular diseases. Using *in vitro* beating primary cultures of murine cardiomyocytes, we have verified the Frank-Starling law of the heart according to the following terms.

In the spheroidal syncytia, where a better 3D distribution of pressure loads was possible in comparison to flat syncytia, an increasing pressure theoretically caused a shortening of the sarcomeres with consequent decreased inotropy, decreased ergotropy, and increased chronotropy (in the 0 ÷ 200 mmHg pressure range, Figures 3, 4, 5, and 6); on the other hand, in the flat syncytia, where the increasing pressure loads uniformly spread onto a 2D surface, the sarcomeres were theoretically stretched with consequent increased inotropy, increased ergotropy, and decreased chronotropy (in the 0 ÷ 100 mmHg pressure range, Figures 3, 4, 5, and 6).

The preceding trends were then inverted for pressures higher than 100 mmHg and 200 mmHg in flat and spheroidal syncytia, respectively. We hypothesize the presence of overstretched sarcomeres in the flat and thin syncytia already at low pressures (>100 mmHg) with consequent impairment of the contractile function. On the other hand, in the spheroidal and thick syncytia, where a better 3D distribution of pressure loads was possible in comparison to the flat and thin syncytia, we hypothesize increasingly packed myofibrils, that is, increasingly cross-linked actin and myosin filaments only at high pressures (>200 mmHg) with consequent amelioration of the contractile function.

In addition, in comparison to the flat and thin syncytia, the spheroidal and thick syncytia showed, at each pressure, higher inotropy, higher ergotropy, and lower chronotropy (Figures 3, 4, 5, and 6): this result was consistent with the general relationship between the contractile parameters and the heart size in mammalians, where the chronotropy decreases due to increasing heart size, whereas the inotropy and the ergotropy increase together with the heart size [21].

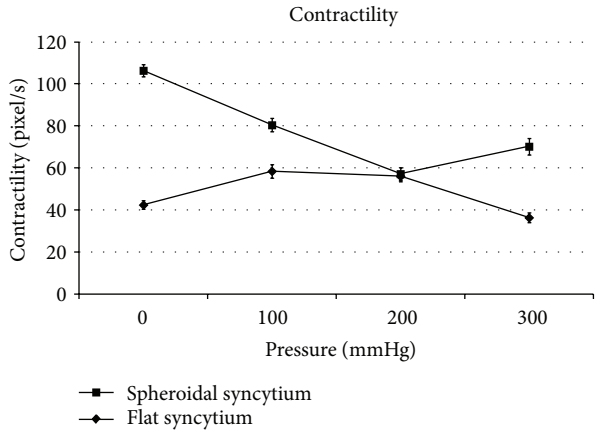

Figure 3: Contractility: the spheroidal syncytia showed minimum contractility at 200 mmHg (with statistically significant differences in comparison with the adjacent values at 100 and 300 mmHg [$P < 0.05$]); the flat syncytia were characterized by maximum contractility at 100 mmHg (with statistically significant difference in comparison with the value at 0 mmHg [$P < 0.05$] and without statistically significant difference in comparison with the value at 200 mmHg [$P > 0.05$]). Except at 200 mmHg, the contractility of spheroidal syncytia was significantly higher than the contractility of flat syncytia ($P < 0.05$).

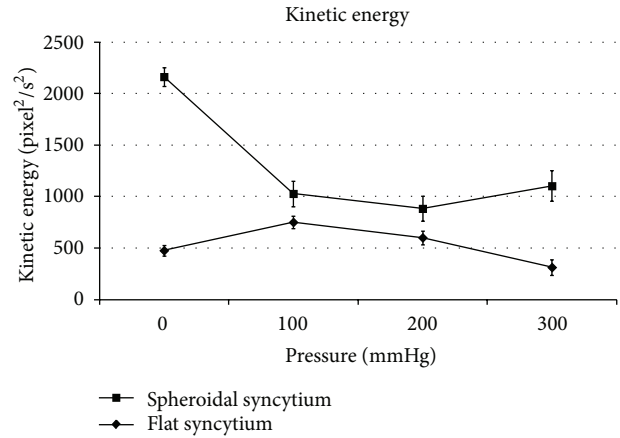

Figure 5: Kinetic energy: the spheroidal syncytia showed minimum kinetic energy at 200 mmHg (without statistically significant differences in comparison with the adjacent values at 100 and 300 mmHg [$P > 0.05$]); the flat syncytia were characterized by maximum kinetic energy at 100 mmHg (with statistically significant differences in comparison with the adjacent values at 0 and 200 mmHg [$P < 0.05$]). At all pressures, the kinetic energy of spheroidal syncytia was significantly higher than the kinetic energy of flat syncytia ($P < 0.05$).

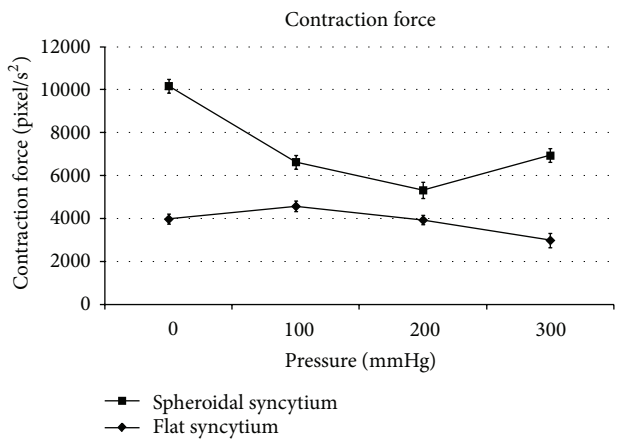

Figure 4: Contraction force: the spheroidal syncytia showed minimum contraction force at 200 mmHg (with statistically significant differences in comparison with the adjacent values at 100 and 300 mmHg [$P < 0.05$]); the flat syncytia were characterized by maximum contraction force at 100 mmHg (with statistically significant differences in comparison with the adjacent values at 0 and 200 mmHg [$P < 0.05$]). At all pressures, the contraction force of spheroidal syncytia was significantly higher than the contraction force of flat syncytia ($P < 0.05$).

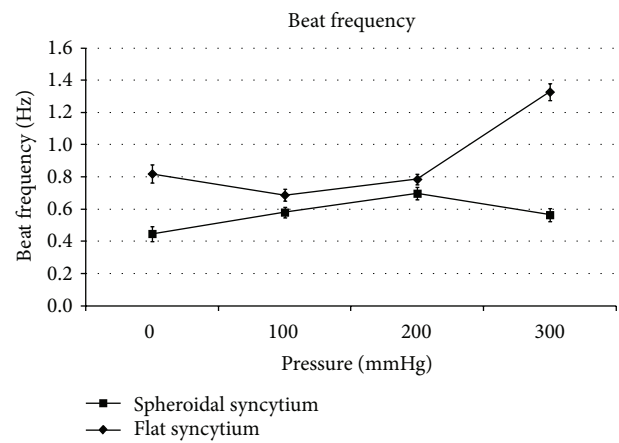

Figure 6: Beat frequency: the spheroidal syncytia showed maximum beat frequency at 200 mmHg (with statistically significant differences in comparison with the adjacent values at 100 and 300 mmHg [$P < 0.05$]); the flat syncytia were characterized by minimum beat frequency at 100 mmHg (with statistically significant differences in comparison with the adjacent values at 0 and 200 mmHg [$P < 0.05$]). At all pressures, the beat frequency of spheroidal syncytia was significantly lower than the beat frequency of flat syncytia ($P < 0.05$).

## 5. Conclusions

In conclusion, the employed calculus method (based on image processing analysis [17]) permitted a systematic study of *in vitro* beating syncytia, which were previously described in terms of cardiac markers and functional gap junctions [22]. As a consequence, it could be used in *in vitro* studies of beating cardiac patches, as alternative to Langendorff's heart in biochemical, pharmacological, and physiology studies, and, especially, when Langendorff's technique is inapplicable (e.g., in studies about human cardiac syncytium in physiological and pathological conditions, patient-tailored therapeutics, and syncytium models derived from induced pluripotent/embryonic stem cells with genetic mutations).

Furthermore, the method could help, in heart tissue engineering and bioartificial heart researches, to "engineer the heart piece by piece" [23]. In particular, our method could be useful in (i) the identification of a suitable cell source, preferably adult-derived autologous stem cells, (ii) the development of biomaterials, and (iii) the design of novel bioreactors and microperfusion systems.

## Appendix

Being both contraction and relaxation active phases of the syncytium movement, in order to estimate a possible ergotropic effect of the pressure, we have defined $E$ as the mean kinetic energy of a beating syncytium in a discrete video:

$$E = \frac{1}{2} A \frac{B}{NM} \sum_{i=1}^{N} \sum_{j=1}^{M} |\underline{v}_{i,j}|^2 \quad \text{in [joule]}, \quad (A.1)$$

where $\underline{v}_{i,j}$ is the velocity of the marker $i$ in the frame $j$, $M$ is the total number of video frames, $N$ is the total number of markers ($N = 30$), $A$ is the constant related to the tissue mass, and $B$ is the constant derived from the linear relation between the units meter and pixel in a bitmap AVI video at a given magnification.

According to (A.1), for each syncytium, in order to compare the pressure effects, there was no need to know the mass of the beating tissue or the $A$ constant, because that mass and that constant were the same in the four different conditions of relative hydrostatic pressure and the spot markers were juxtaposed in the same grid positions. In addition, there was no need to know the video metrics or the $B$ constant, because the metrics and constant and the video magnification were the same at all pressures. As a consequence, we have defined $E_{\text{norm}}$ as the normalized mean kinetic energy of a beating syncytium in a discrete video:

$$E_{\text{norm}} = \frac{E}{(1/2) AB} = \frac{1}{NM} \sum_{i=1}^{N} \sum_{j=1}^{M} |\underline{v}_{i,j}|^2 \quad \text{in [pixel}^2/\text{s}^2]. \quad (A.2)$$

Besides, for each marker $i$, in order to calculate $B_i$ as the total number of beats, we have identified and counted the peak displacements. Given the duration $T$ of the video ($T = 40$ s), we have defined $f$ as the mean beat frequency:

$$f = \frac{1}{TN} \sum_{i=1}^{N} B_i \quad \text{in [Hz]}. \quad (A.3)$$

According to Sonnenblick et al. [24, 25], the isotonic contraction of the papillary tissue is well described by the force-velocity curves, where, at a given load, the maximum contraction velocity is an indicator of contractility.

As a consequence, in order to study a possible inotropic effect of the pressure under a kinematic point of view, for each marker $i$ during its $B_i$ beats, we have identified the peak velocities and we have evaluated the mean syncytium contractility $C_{\text{mean}}$ as their mean. $C_{\text{mean}}$ is expressed in [pixel/s].

In order to study a possible inotropic effect of the pressure under a dynamic point of view, we have evaluated the syncytium contraction force by the Hamiltonian mechanics. The so-called Hamiltonian function $H$ is the sum of the kinetic and potential energy. Assuming that, during the whole video observation, there was a plentiful source of available glucose from the culture medium and that the subsequent ATP production and distribution were isotropic, $P_{\text{ATP}}$, the ATP-related potential energy for the contraction movement, could be supposed to be constant in time and in space. As a consequence, the Hamilton differential equations to describe the syncytium movement were:

$$F_x = -\frac{\partial H}{\partial x} = -\frac{\partial}{\partial x}(E_{\text{ATP}} + P_{\text{ATP}}) = -\frac{\partial E_{\text{ATP}}}{\partial x}$$

$$\text{in [newton]}$$

$$F_y = -\frac{\partial H}{\partial y} = -\frac{\partial}{\partial y}(E_{\text{ATP}} + P_{\text{ATP}}) = -\frac{\partial E_{\text{ATP}}}{\partial y}$$

$$\text{in [newton]}, \quad (A.4)$$

where $F_x$ and $F_y$ are the orthogonal components of the contraction force $\underline{F}$ and $E_{\text{ATP}} = E_{\text{ATP}}(x, y, t)$ is the kinetic energy function of the beating syncytium.

From (A.4), we have obtained:

$$F_x = -\frac{\partial}{\partial x}\left(\frac{1}{2}mv^2\right) = -\frac{1}{2}m\frac{\partial}{\partial x}(v_x^2 + v_y^2)$$

$$= -m\left(v_x \frac{\partial v_x}{\partial x} + v_y \frac{\partial v_y}{\partial x}\right)$$

$$F_y = -\frac{\partial}{\partial y}\left(\frac{1}{2}mv^2\right) = -\frac{1}{2}m\frac{\partial}{\partial y}(v_x^2 + v_y^2) \quad (A.5)$$

$$= -m\left(v_x \frac{\partial v_x}{\partial y} + v_y \frac{\partial v_y}{\partial y}\right),$$

where $m$ is the beating tissue mass and $v_x$ and $v_y$ are the orthogonal components of the velocity vector $\underline{v}$.

Combining (A.5), we have obtained:

$$\underline{F} = \begin{pmatrix} F_x \\ F_y \end{pmatrix} = -m \begin{pmatrix} \dfrac{\partial v_x}{\partial x} & \dfrac{\partial v_y}{\partial x} \\ \dfrac{\partial v_x}{\partial y} & \dfrac{\partial v_y}{\partial y} \end{pmatrix} \begin{pmatrix} v_x \\ v_y \end{pmatrix}$$

$$= -m \begin{pmatrix} \dfrac{\partial v_x}{\partial x} & \dfrac{\partial v_y}{\partial x} \\ \dfrac{\partial v_x}{\partial y} & \dfrac{\partial v_y}{\partial y} \end{pmatrix} \underline{v}. \quad (A.6)$$

According to (A.6), in order to study a possible inotropic effect of the pressure under a dynamic point of view, we have defined $F_{\text{mean}}$ as the normalized mean contraction force of a beating syncytium in a discrete video:

$$F_{\text{mean}} = \frac{1}{AB} \frac{1}{NM} \sum_{i=1}^{N} \sum_{j=1}^{M} \left| \underline{F}_{i,j} \right| \quad \text{in } \left[\text{pixel/s}^2\right]. \quad (A.7)$$

According to Cauchy's theory of continuum mechanics, introducing the $X$ and $Y$ reference axes of the undeformed tissue configuration, we could rewrite (A.6) as follows:

$$\begin{aligned}
\underline{F} &= -m \frac{\partial}{\partial t} \left[ \begin{pmatrix} \frac{\partial x}{\partial X} & \frac{\partial y}{\partial X} \\ \frac{\partial x}{\partial Y} & \frac{\partial y}{\partial Y} \end{pmatrix} \right] \underline{v} \\
&= -m \frac{\partial}{\partial t} \left[ \begin{pmatrix} \frac{\partial x}{\partial X} & \frac{\partial x}{\partial Y} \\ \frac{\partial y}{\partial X} & \frac{\partial y}{\partial Y} \end{pmatrix}^T \right] \underline{v} = -m \frac{\partial}{\partial t} \left[ (\nabla \underline{s})^T \right] \underline{v}.
\end{aligned} \quad (A.8)$$

In (A.8), under the hypothesis of small displacements, the symmetric part of the displacement gradient ($\nabla \underline{s}$) is the deformation tensor, whereas the nonsymmetric one describes the rigid movement.

In our experimental setup, (A.8) was then suitable for studying the contraction force both in terms of pure deformation (symmetric part of the displacement gradient) and in terms of roto-translation (nonsymmetric part of the displacement gradient).

## Ethical Approval

All procedures involving mice were completed in accordance with the policy of the Italian National Institute of Health (Protocol no. 118/99-A) and with the ethical guidelines for animal care of the European Community Council (Directive no. 86/609/ECC). CD-1 mice were obtained from the Charles River Laboratories Italia (Calco, Italy) and were housed under 12 h light/dark cycles, at constant temperature, and with food and water *ad libitum*. The mice were sacrificed by $CO_2$ asphyxiation.

## Conflict of Interests

The authors declare that there is no conflict of interests regarding the publication of this paper.

## Acknowledgments

This work was supported by the following research projects: FIRB Grant 2010 and INAIL Grant 2010 to Fabio Naro.